\documentclass[namedreferences]{solarphysics}
\usepackage[hyperref,optionalrh]{spr-sola-addons}
\usepackage{graphicx} 
\usepackage{color}        
\usepackage{breakurl}

\chardef\us=`\_

\usepackage{savesym}
\savesymbol{iint}
\savesymbol{iiint}
\usepackage{amsmath}
\usepackage{tabularx}
\usepackage{amsfonts}
\usepackage{booktabs}
\restoresymbol{TXF}{iint}
\restoresymbol{TXF}{iiint}

\begin{document}
\begin{article}
\begin{opening}
\title{Stokes inversion techniques with neural networks: analysis of uncertainty in parameter estimation}

\runningauthor{Lukia Mistryukova}
\runningtitle{Solar atmosphere parameters uncertainty}

\author[addressref={aff1}, email={lmistryukova@hse.ru
}]{\inits{L.}\fnm{Lukia}~\lnm{Mistryukova}}

\author[addressref={aff2,aff3},email={}]{\inits{A.}\fnm{Andrey}~\lnm{Plotnikov}}

\author[addressref={aff1},email={}]{\inits{I.}\fnm{Aleksandr}~\lnm{Khizhik}}

\author[addressref={aff2,aff4},email={}]
{\inits{I.}\fnm{Irina}~\lnm{Knyazeva}}

\author[addressref={aff1},email={}]
{\inits{I.}\fnm{Mikhail}~\lnm{Hushchyn}}

\author[addressref={aff1},email={}]
{\inits{I.}\fnm{Denis}~\lnm{Derkach}}

\address[id={aff1}]{HSE University, Moscow, Russia}
\address[id={aff2}]{Pulkovo Observatory, Saint-Petersburg, Russia}
\address[id={aff3}]{Crimean Astrophysical Observatory, Crimea}
\address[id={aff4}]{Saint-Petersburg State University, Saint-Petersburg, Russia}

\begin{abstract}
Magnetic fields are responsible for a multitude of Solar phenomena, including such destructive events as solar flares and coronal mass ejections, with the number of such events rising as we approach the peak of the 11-year solar cycle, in approximately 2025. High-precision spectropolarimetric observations are necessary to understand the variability of the Sun. The field of quantitative inference of magnetic field vectors and related solar atmospheric parameters from such observations has long been investigated. In recent years, very sophisticated codes for spectropolarimetric observations have been developed. Over the past two decades, neural networks have been shown to be a fast and accurate alternative to classic inversion technique methods. However, most of these codes can be used to obtain point estimates of the parameters, so ambiguities, the degeneracies, and the uncertainties of each parameter remain uncovered. In this paper, we provide end-to-end inversion codes based on the simple Milne-Eddington model of the stellar atmosphere and deep neural networks to both parameter estimation and their uncertainty intervals. The proposed framework is designed in such a way that it can be expanded and adapted to other atmospheric models or combinations of them. Additional information can also be incorporated directly into the model. It is demonstrated that the proposed architecture provides high accuracy of results, including a reliable uncertainty estimation, even in the multidimensional case. The models are tested using simulation and real data samples.
\end{abstract}

\end{opening}

\section{Introduction}
Modern solar physics relies, to a great extent, on the spectropolarimetric observations of the Sun~\citep{ramos2016inversion, gafeira2021machine, li2022tic, leka2022on}. The data on spectral and polarization state of the solar light, in conjunction with an appropriate atmosphere model allows one to derive the thermodynamic, dynamic, and magnetic properties of the solar plasma~\citep{viticchie2011asymmetries}. Over the past two decades, many approaches aimed at deriving the solar atmosphere parameters have been developed. Despite great capabilities provided by these techniques, in many cases, the derivation of the atmospheric parameters from the observed spectra -- the inverse problem -- requires huge computing resources.

The formation of the spectral line in the solar atmosphere is described by the radiation transfer equation~\citep[RTE,][]{degl2004polarization}. In turn, the polarization of light, which appears due to Zeeman splitting of the spectral line in an external magnetic field, can be described by four components I, Q, U, and V of the observed Stokes vector~\citep{kuckein2021multiple}.

In the general case, the complexity of the RTE makes it impossible to solve the inverse problem analytically, that is, to obtain the solar atmosphere parameters from the observed Stokes profiles. Certain simplifications and approximations are often used to solve the RTE, for instance, the Milne-Eddington (ME) model of the atmosphere~\citep{unno1956line, degl2004polarization, DelToroIniesta2016}. The model assumes a local thermal equilibrium and independence of the atmosphere parameters with height. Nevertheless, in such a case, the solution of the RTE is still non-linear and transcendent. 

To fill the gap between theoretical models and complex simulations, special tools called inversion codes are developed. These codes are classified according to the optimization strategies used to find the optimal set of atmospheric parameters that generate spectra closest to the observed ones~\citep{lites2007a}. In most cases the procedure is based on non-linear least squares optimization with the simplified atmospheric model. More complicated models exist, that take into account height-dependent distribution of atmospheric parameters (non-local thermal equilibrium).

Data for training, validation and testing contain spectropolarimetric images where each pixel of the image corresponds to an area on the Sun’s surface with its spectral profile. These images are converted into a three-dimensional data set $(x, y, \lambda)$, in which $\lambda$ is the width of the pixel spectral line with coordinates $(x, y)$. In this context, a data dimensionality problem arises: each pixel of an image is equal to one independent inversion problem. This makes the methods based on optimization techniques computationally expensive. As an example, to analyze data from the Helioseismic and Magnetic Imager (HMI) on-board of the Solar Dynamics Observatory (SDO) using Very Fast Inversion of the Stokes Vector~\citep[VFISV,][]{borrero2011vfisv}, 50 CPUs were used in parallel to reach a 10-minute cadence, including specific limitations, such as additional assumptions for the atmospheric model and low spectral resolution. Also, the result depends on the closeness of the initial approximation to the true value of the parameters.
Furthermore, the atmospheric model itself is a multi-dimensional problem. In the simple basic ME model, there is a set of 11 atmospheric parameters which form a manifold in the parameters space. As a result, the inversion task became ill-conditioned and it may lead to instability of estimations or even multiple solutions. That is why a proper estimation of the uncertainty becomes an important issue.

As a possible solution to overcome the computational problems, a neural-network based solution was proposed~\citep{carroll2001the}. The main idea of this approach is to use neural networks for direct inversion by studying the mapping between Stokes profiles and atmospheric parameters. In the multiple studies this approach has shown to be effective, however in most cases the uncertainty estimation problem still need to be solved. This is connected to a large number of parameters the neural solutions operate, which requires a lot of computing power to scan~\citep{ghahramani2015probabilistic, krzywinski2013power}. In the following text, we provide a review of main studies connected with a neural network application to Stocks inversion problems and current state of uncertainties estimation. The list with a short description and links to the paper known to us related to the Stokes inversion problem is summarized in Tab.~(\ref{review}).

In one of the first papers~\citep{carroll2001the}, several neural networks were used for parameters recovering from the atmosphere model with temperature stratification. DIAMAG synthesis codes~\citep{grossmann1994thedeep}, on the basis of the nine semi-empirical model atmospheres, with temperature and pressure stratification, were used to generate an input database consisting of I, Q, U and V profiles. As a result, for each observed profile, the probability that this particular model has been responsible for producing it was calculated. Next, the MLP network was trained for each of the nine models to recover several atmospheric parameters, including three components of magnetic field vector, velocities and turbulence.

Ramos and Baso (\citeyear{ramos2019stokes}) developed inversion codes of atmospheric physical properties, based on a fully convolutional neural network, trained on synthetic Stocks profiles generated with the state-of-the-art MPS/University of Chicago Radiative (MURaM) MHD codes, which are three-dimensional magneto-hydro\-dynamic numerical simulations of different structures of the solar atmosphere. The authors have shown that this approach gives results comparable to those of standard inversion techniques. Gafeira et al. (\citeyear{gafeira2021machine}) pointed out that neural network based methods failed to take into account the physical connections between the parameters and physical model and suggested using neural networks, not as a tool to assist inversion, but precisely as an initial guess for further Stokes profile inversions.

Another approach, based on real data, was described in~\citep{guo2021a} or~\citep{liu2020inferring}. The authors of the latter used real data taken from the Near InfraRed Imaging Spectropolarimeter at the Big Bear Solar Observatory. Standard ME inversion codes were used for data labeling; after that, the convolution neural networks were trained to match the input spectrum and inverted with the ME codes parameters. The authors show that neural networks learn to approximate ME inversion results very closely. The main result of this paper was in computational efficiency compared to ME codes.
Two-dimensional convolutions can increase the potential of a model by taking into account spatial correlations in Stokes parameters and correlations along the optical beam. As a result, the model can provide a three-dimensional cube of parameters of the studied region, taking into account the geometric height, and the ability to recover parameters that cannot be recovered using the classic approaches.

The decline in the performance of neural network models trained on synthetic spectra is demonstrated in the article~\citep{socas-navarro2005strategies} for the task of restoring magnetic field strength. Two approaches are proposed in this article to address this deficiency. Preprocessing of the observed profiles in order to project them onto ME hyperspace allows the efficient finding of ME profiles closest to the observed ones, which are the input of the neural network model. As such, the article uses Auto-associative Neural Networks (AANNs) -- models that have at least one intermediate layer with fewer neurons than the input and output layers. The description of other methods of preprocessing, such as PCA and Expansion in Hermitian Functions (EHF) and their comparison with AANN can be found in~\citep{socas-navarro2005strategies}. The second approach is to regularize the model, in order to weaken its attention to small deviations from the ideal form, which have unprocessed synthetic profiles. To do this, artificial perturbations can be added to the profiles to violate their symmetry.

Several architectures containing multi-layer perceptron (MLP) blocks were compared in the paper~\citep{knyazeva2022multi} using data with synthetic spectra. Three models were considered: one with individual MLP blocks to recover each parameter separately, one with a common MLP model for all parameters and a partly sharing model consisting of a common MLP block and independent MLP blocks, each of which corresponded to a certain parameter. It was found that the partly sharing model shows the best performance. 

Within the last several years, various approaches have been developed to estimate the prediction uncertainty. These methods can be split into four groups, based on the number and the nature of the used neural networks~\citep{lakshminarayanan2017simple}: \textit{single deterministic methods}, where the uncertainty on a prediction is computed based on one single forward pass within a deterministic network~\citep{malinin2018predictive}, \textit{Bayesian methods} that contain uncertainty in their networks, assuming that parameters are defined as some probability distributions~\citep{blundell2015weight}, \textit{augmentation methods}, which are based on the modification of the training data set, so a model learns on the extended data~\citep{shorten2019a}, and \textit{ensembles}, which derive a final prediction based on other predictions received from multiple ensemble members~\citep{lakshminarayanan2017simple}.

\begin{figure*}
\centering
\includegraphics[width=1\textwidth]{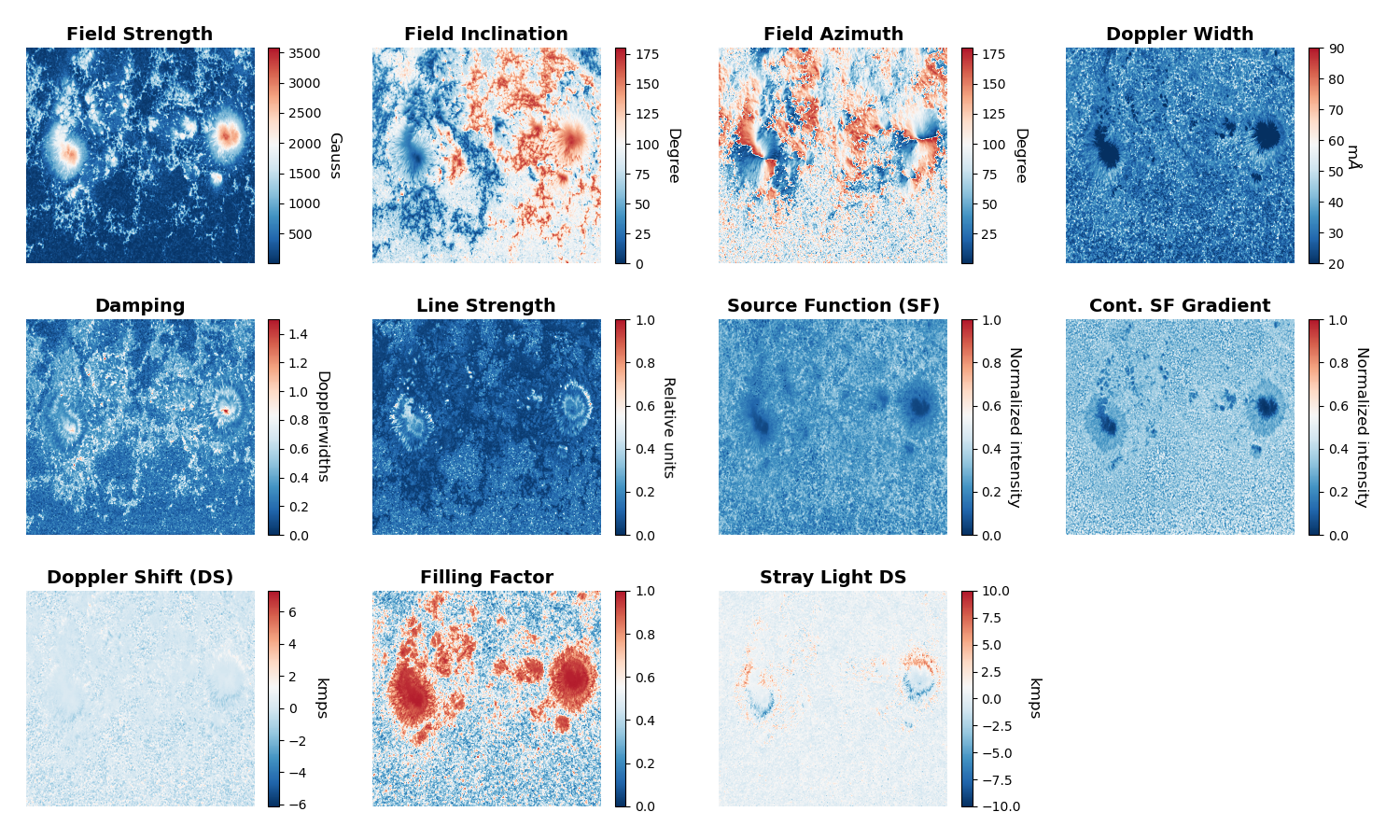}
\caption{Two-dimensional color maps of Stokes profiles inversion by Milne-Eddington codes HAO "MERLIN" (reference values).}
\label{true_params}
\end{figure*}

One of the probabilistic approaches to solving the inverse problem is to use algorithms based on the Bayesian inference, such as the Markov Chain Monte Carlo (MCMC) and Nested Sampling (NS). Despite limitations in their applicability, even for relatively simple atmospheric models and requirements of detailed knowledge of the parameter space, Bayesian approaches have long been used in solar physics analysis~\citep{li2019mcmc, ramos2007bayesian}. An alternative to these may be variational inference methods, for example, normalizing flows, where the true distribution of the solution is approximated by a simpler analytical one~\citep{baso2021bayesian, ramos2017inference}. However, the disadvantage of such models is that their optimization is not always stable, so one has to consider the more simple, posterior distribution.

Recently, it has been proposed to use a single deterministic network to quantify uncertainties in the prediction of atmospheric parameters~\citep{higgins2021fast, higgins2022synthia}. The authors of these papers suggest to obtain confidence intervals treating the problem as a regression by classification, so the model predicts for each pixel of the parameter image the distribution of its possible values by applying the softmax function.

In this paper, we focus on the uncertainty estimation using a combination of several convolutional neural networks, modified in such a way that they can quantify the uncertainty of predictions (treating the observed value as a sample from a Gaussian distribution) and trained as an ensemble. 
To the best of our knowledge, this is the first systematic study of these approaches within the framework of the inverse solar problem.

\section{Uncertainty Quantification and Metrics}\label{sect:coverage}
Experimental measurement uncertainty plays a central role in physical sciences. The assigned uncertainties can point to the reliability of the measurement. That is why, currently, the interpretation of uncertainty plays a crucial role in the analysis of the experiment. In this paper, we follow the most frequent interpretation, which leads to an important consequence: the methods proposed should estimate interval with a given confidence, that is provide experimental coverage probability for a given confidence level. This in turn means obtaining the rate at which the true value is contained in the confidence interval of an individual measurement~\citep{pawitan2001in}.

\begin{figure}
\centering
\includegraphics[scale=0.45]{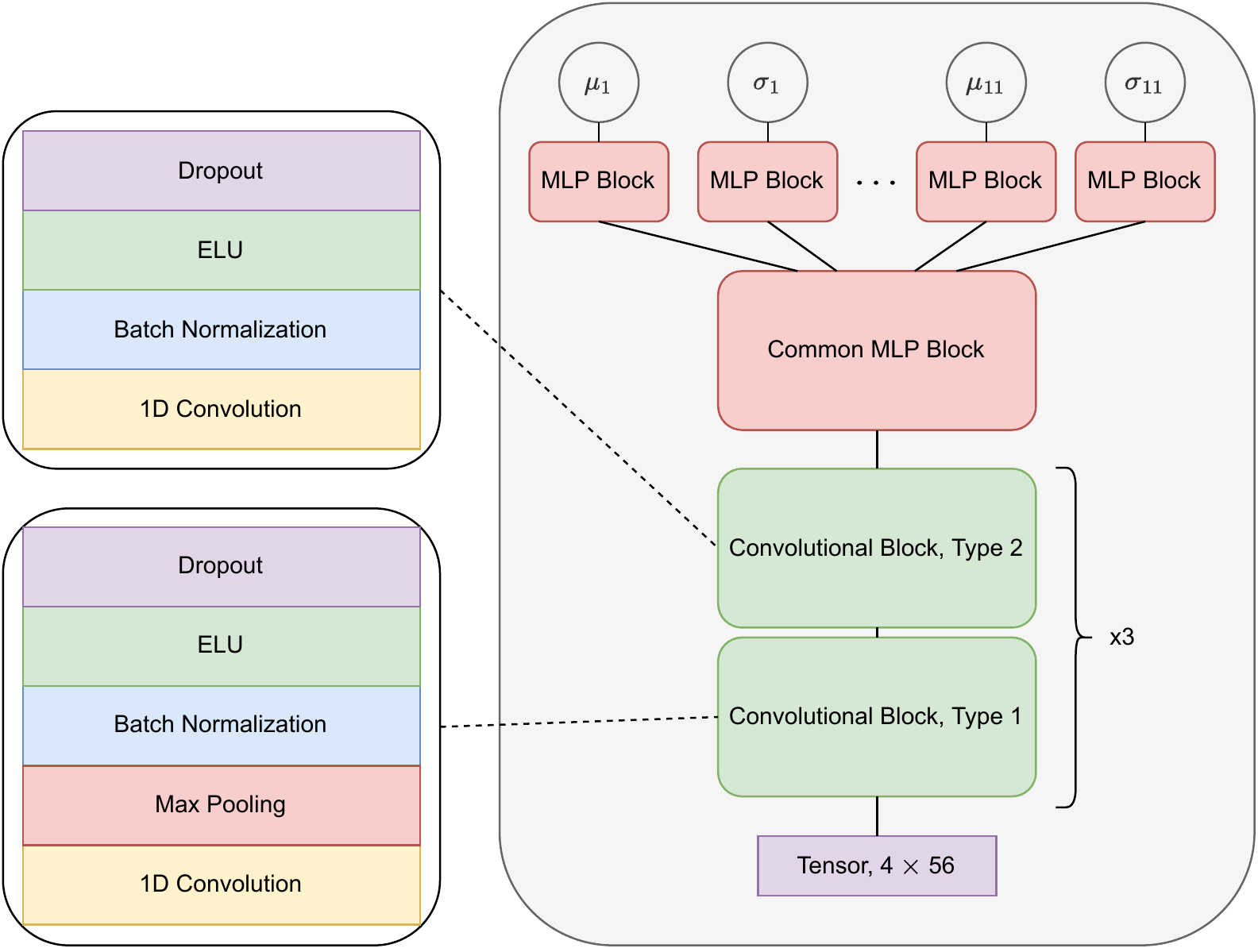}
\caption{Network architecture used in atmospheric parameters inferring.
The model has 22 independent MLP blocks, corresponding to the mean and the variance of 11 atmospheric parameters, one common MLP block and 6 common convolutional blocks of two types. Blocks of the second type contain the following layers: one-dimensional convolution, batch normalization, ELU activation function and dropout. In blocks of the first type the max pooling layer is added.}
\label{conv_architecture}
\end{figure}

We check the correctness of the procedure using a graphical representation as shown later in Figs.~(\ref{unc_1}) and (\ref{unc_2}). We also use integral metrics to estimate how accurately we evaluate the confidence of our model: the normalized Mean Squared Error (nMSE)~\citep{quinonerocandela2005evaluating}, the Negative Log Predictive Density (NLPD)~\citep{quinonerocandela2005evaluating} and the Prediction Interval Coverage Probability (PICP)~\citep{shrestha2006machine} with two different fractions of the distribution inside the confidence interval. 
The nMSE is defined as
\begin{equation} \label{nMSE}
\text{nMSE} = \sqrt{\frac{1}{N} \sum_{i=1}^{N} \frac{\left(t_{i}-m_{i}\right)^{2}}{\sigma_i^2}},
\end{equation}
where $m_i$, $\sigma_i^2$ are the mean and variance of the predictive distribution, respectively, $N$ denotes the total number of pixels (the size of a test data set) and $t_i$ is a true sample. The NLPD is defined as
\begin{equation} \label{NLPD}
 \text{NLPD} = \frac{1}{N} \sum_{i=1}^{N} \left( \frac{\left(t_{i}-m_{i}\right)^{2}}{2 \sigma_i^2} + \log{\sigma_i} + c \right),
\end{equation}
where $c$ is a constant independent of $m_i$ and $\sigma_i$. 
Specifying the upper $\text{PL}_{i}^{U}$ and lower $\text{PL}_{i}^{L}$ bounds on a prediction $i$ (the uncertainty on a single observation) one can calculate the prediction interval coverage probability metric (PICP). The PICP is defined as the probability that the real value lies within the predicted confidence interval and estimated as a following frequency:
\begin{equation} \label{PICP}
\text{PICP}=\frac{1}{N} \sum_{i=1}^{N} m_i, ~~ m_i = \begin{cases}
1, & \text{PL}_{i}^{L} \leq t_{i} \leq \text{PL}_{i}^{U}\\
0, & \text{otherwise}
\end{cases},
\end{equation}
where $t_i$ is the $i$-th element of the reference data set. Both the nMSE and PICP metrics most penalize the cases of incorrect forecasts made with uncertainty close to zero. However, PICP is sensitive both when the prognosis is not sufficiently certain and when it is over-confident. The PICP value should be close to the $\alpha$, however over-confident predictions are worse for the model. 

These metrics show the overall performance of the algorithm, while the local performance might vary depending on the point in the parameter space.

\begin{figure*}
\centering
\includegraphics[width=1\textwidth]{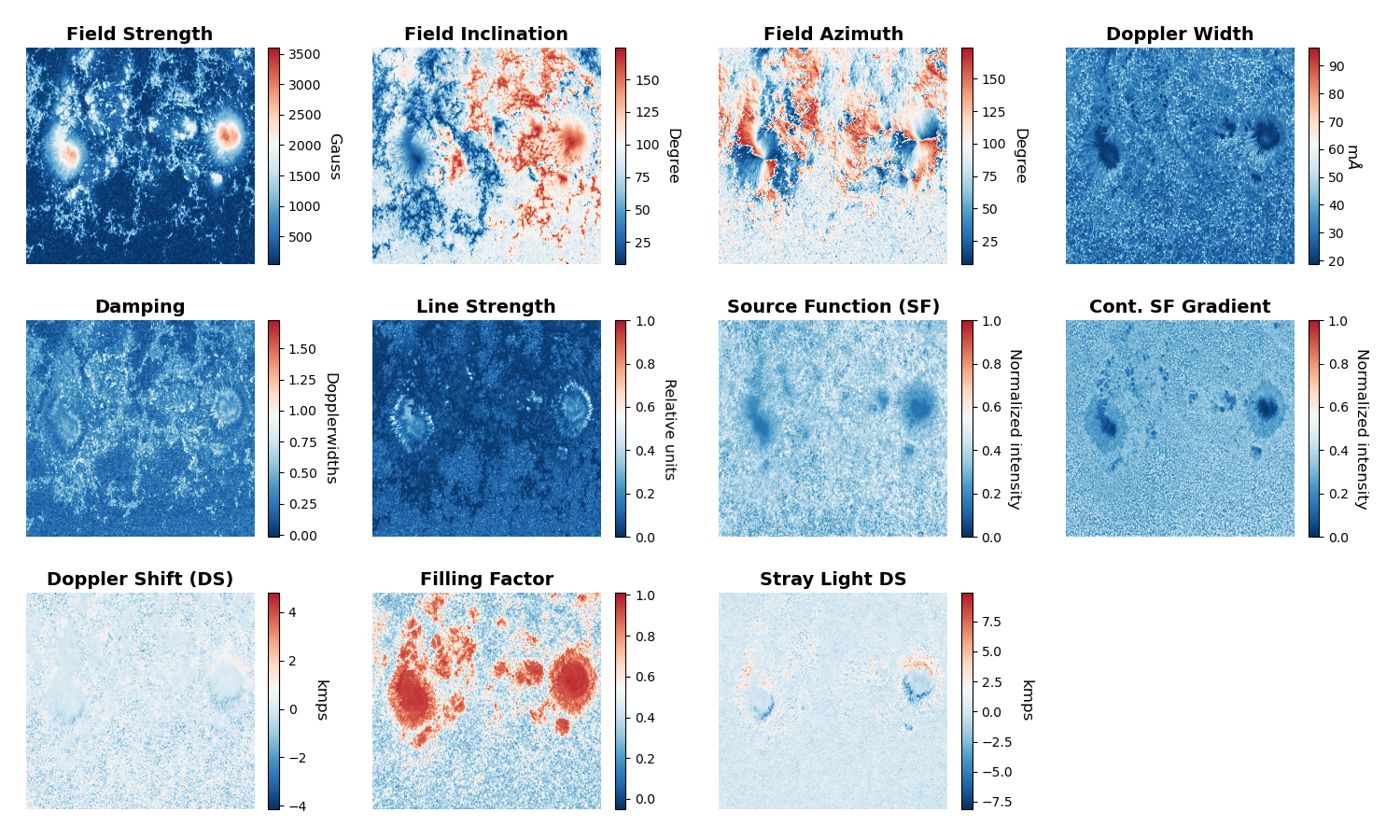}
\caption{Two-dimensional color maps of parameters predicted by the model.}
\label{params}
\end{figure*}

\section{Data and methods} \label{sect:data_and_methods}
The solution of the inverse problem requires, at the first stage, the initial approximation of the Stokes profiles. These profiles can be generated synthetically using atmospheric models, and this method has been shown to be highly effective~\citep{knyazeva2022multi}. ME codes obtained from the \href{https://www2.hao.ucar.edu/csac}{Hinode/SOT/SP database} were used to collect 5.3 million synthetic records of the state of the solar atmosphere. Generated spectra were made more similar to real samples by adding Gaussian noise, and then were used for training. 

In the following study, we consider 11 parameters of the solar atmosphere: the magnetic field vector, consisting of field strength component, inclination angle and azimuth angle; line parameters, consisting of Doppler width, line damping and line strength; two intensity parameters, consisting of source function (SF) and its height gradient; Doppler shift (DS) of the line; stray light Doppler shift and magnetic filling factor. Before training, true samples of parameter values were transformed by applying logarithmic transformation to magnetic field vector and trigonometric to magnetic field inclination and azimuth angles and were brought to one scale with min-max normalization. The example of color maps is shown in Fig.~(\ref{true_params}).

\begin{table}
\begin{tabularx}{\textwidth}{lccccccc}
\toprule
\textbf{Parameter} & \textbf{R}$^2$ & \textbf{MSE} & \textbf{MAE} & \textbf{NLPD} & \textbf{nRMSE} & \textbf{PICP}$_{68}$ & \textbf{PICP}$_{95}$ \\ 
\midrule
Field Strength & 0.914 & 0.0015 & 0.019 & -2.735 & 0.977 & 0.792 & 0.959 \\ 
Field Inclination & 0.952 & 0.0034 & 0.025 & -2.356 & 1.079 & 0.808 & 0.959 \\ 
Field Azimuth & 0.735 & 0.0313 & 0.093 & -1.356 & 0.721 & 0.866 & 0.982 \\ 
Doppler Width & 0.970 & 0.0005 & 0.015 & -2.676 & 0.874 & 0.756 & 0.977 \\ 
Damping & 0.934 & 0.0015 & 0.019 & -2.658 & 0.869 & 0.770 & 0.974 \\ 
Line Strength & 0.821 & 0.0025 & 0.019 & -2.790 & 0.913 & 0.795 & 0.968 \\ 
SF & 0.903 & 0.0005 & 0.013 & -2.769 & 0.848 & 0.781 & 0.978 \\ 
Cont. SF Grad. & 0.981 & 0.0002 & 0.009 & -3.195 & 0.819 & 0.797 & 0.982 \\ 
DS & 0.956 & 0.0001 & 0.004 & -4.220 & 0.779 & 0.820 & 0.984 \\ 
Filling Factor & 0.859 & 0.0096 & 0.062 & -1.345 & 0.963 & 0.754 & 0.954 \\ 
Stray Light DS & 0.639 & 0.0023 & 0.014 & -1.875 & 1.773 & 0.858 & 0.973 \\
\bottomrule
\end{tabularx}
\caption{Performance metrics of the convolutional model.}
\label{p_metrics}
\end{table}

We used a single deterministic method to evaluate the predictive uncertainty and modified architecture of the partial sharing model~\citep{knyazeva2022multi} in such way that it predicted two values in the final layer for each pixel, corresponding to the mean $\mu(\mathbf{x})$ and the variance $\sigma^2(\mathbf{x}) > 0$. The proposed model consists of 22 independent MLP blocks, corresponding to the mean value and the standard deviation of each pixel in case of 11 atmospheric parameters. Independent blocks took on the input the result of one common MLP block, which in turn was after 6 common convolutional blocks of two types. Blocks of the first type contain the following layers: one-dimensional convolution with kernel size 3, max pooling with kernel size 2, batch normalization, ELU activation function and dropout. In blocks of the second type the max pooling layer is excluded. It was reasonable to add convolutional blocks to the model, since Stokes profiles are usually interconnected. The schematic representation of the architecture is shown in Fig.~(\ref{conv_architecture}). 

Treating the observed value as a sample from a Gaussian distribution the model was trained by minimizing the negative logarithm loss function~\citep{gawlikowski2021a}:
\begin{equation}\label{NLL}
 L(\theta, y, \mathbf{x}) =- \frac{1}{MK} \sum_{i=1}^{M} \sum_{j=1}^{K} \log \left(\frac{1}{\sqrt{2 \pi} \sigma_{\theta ij}(\mathbf{x})} \exp \left[-\frac{(y_{ij}-\mu_{\theta ij}(\mathbf{x}))^{2}}{2 \sigma_{\theta ij}^{2}(\mathbf{x})}\right]\right),
\end{equation}
where $\mu_{\theta}(\mathbf{x})$ and $\sigma_{\theta}(\mathbf{x})$ -- predicted average value and standard deviation, respectively, $K=11$ and $M=128$ are the number of parameters that were considered and the batch size, respectively. The Adam optimizer was used for the iterative update of the model weights. The model was found to achieve the required result in 5 epochs.

\begin{figure*}
\centering
\includegraphics[width=1\textwidth]{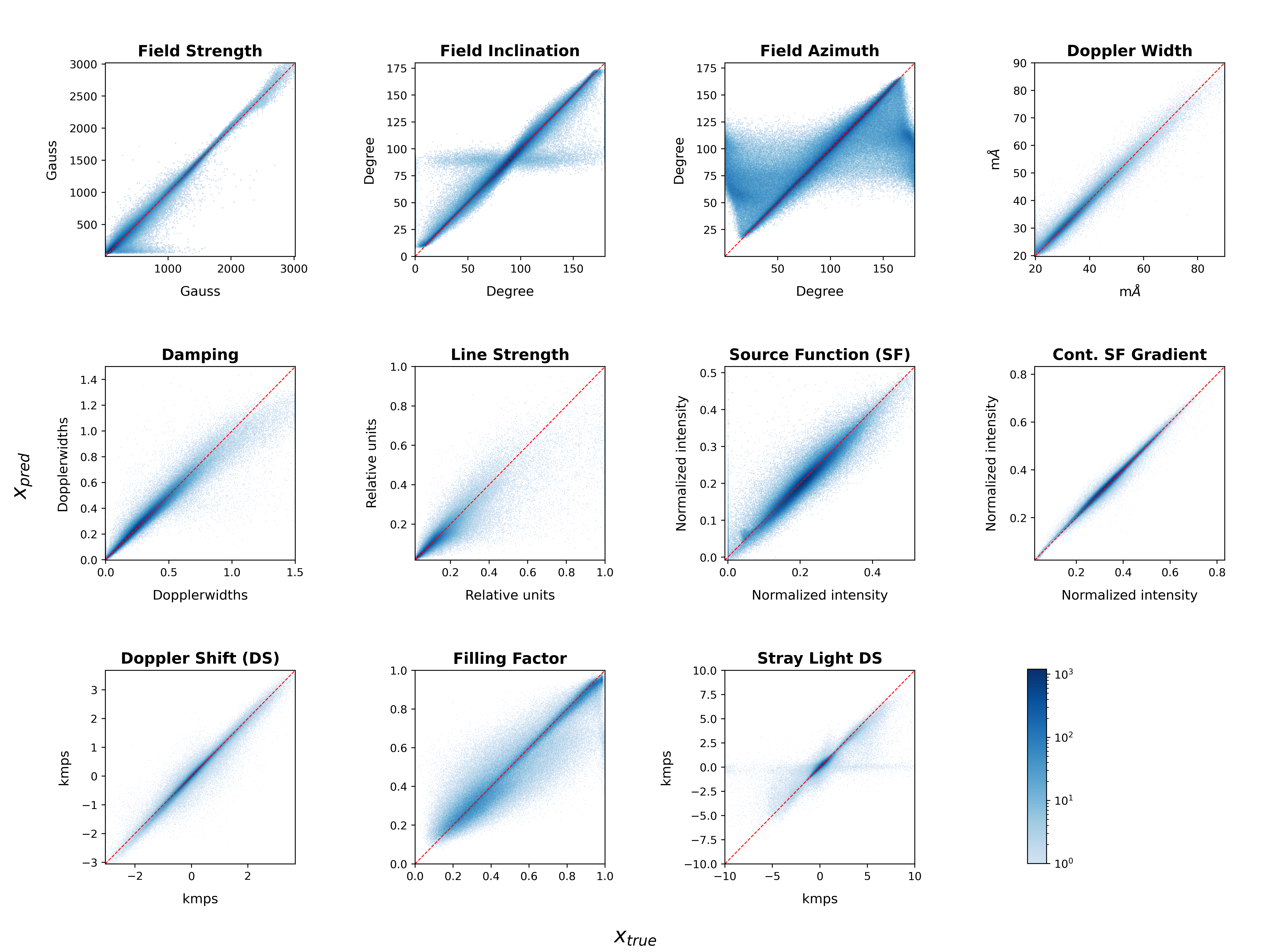}
\caption{Dependence of predictions on the true values. Each figure corresponds to one of the atmospheric parameters. Red dashed lines represent pixels whose predictions are equal to the true values. Color intensity indicates the frequency of the values encountered.}
\label{comparison}
\end{figure*}

While other methods exist~\citep{gawlikowski2021a}, in order to obtain an estimate a scan of the loss-function can be used. This requires a lot of computing resources, is why approximate methods are implemented. These methods allows for the estimate of the minimum width, however, their performance depends on the problem. We test the ensemble-based model~\citep{lakshminarayanan2017simple}, which is shown to perform well in open datasets. The model described above was used to build an ensemble of several models, that were trained independently on different training sub-samples. The final prediction was treated as a uniformly weighted mixture of Gaussian distributions, and the combination of results was determined as follows:
\begin{equation} \label{Ensembles}
\begin{aligned}
 \mu_{*}(\mathbf{x})&=\frac{1}{N} \sum_{i=1}^{N} \mu_{\theta i}(\mathbf{x}),\\ 
 \sigma_{*}^{2}(\mathbf{x})&=\frac{1}{N} \sum_{i=1}^{N}\left(\sigma_{\theta i}^{2}(\mathbf{x})+\mu_{\theta i}^{2}(\mathbf{x})\right)-\mu_{*}^{2}(\mathbf{x}), 
\end{aligned}
\end{equation}
where $\mu_{\theta i}(\mathbf{x})$, $\sigma_{\theta i}(\mathbf{x})$ are the mean and the standard deviation of the $i$-th model, respectively, and $N = 6$ is the number of models used. Predicted values were scaled back into their physical ranges after training to correctly interpret the results. 

\label{sect:results}
\section{Results} 
For each parameter, we compared its ME-calculated values with our network inferred values and computed performance metrics. Three quality metrics are the following: coefficient of determination R$^2$, the mean squared error (MSE) and the mean absolute error (MAE). Additionally, we show the metrics that characterize the uncertainty region, as defined in Sec.~\ref{sect:coverage}. 

\begin{table*}
\begin{tabularx}{\textwidth}{lccccccc}
\toprule
\textbf{Parameter} & \textbf{R}$^2$ & \textbf{MSE} & \textbf{MAE} & \textbf{NLPD} & \textbf{nRMSE} & \textbf{PICP}$_{68}$ & \textbf{PICP}$_{95}$ \\ \midrule
Field Strength & 0.957 & 0.0007 & 0.015 & -2.953 & 1.005 & 0.692 & 0.957 \\ 
Field Inclination & 0.986 & 0.0010 & 0.017 & -2.740 & 1.002 & 0.706 & 0.959 \\ 
Field Azimuth & 0.832 & 0.0188 & 0.064 & -2.013 & 0.925 & 0.752 & 0.965 \\ 
Doppler Width & 0.976 & 0.0004 & 0.014 & -2.764 & 1.009 & 0.683 & 0.953 \\
Damping & 0.967 & 0.0005 & 0.014 & -2.820 & 1.010 & 0.687 & 0.951 \\ 
Line Strength & 0.874 & 0.0009 & 0.014 & -3.003 & 1.038 & 0.708 & 0.948 \\
SF & 0.911 & 0.0004 & 0.013 & -2.834 & 0.965 & 0.706 & 0.962 \\ 
Cont. SF Grad. & 0.983 & 0.0001 & 0.008 & -3.265 & 0.953 & 0.710 & 0.964 \\ 
DS & 0.970 & 0.0001 & 0.004 & -4.305 & 0.959 & 0.710 & 0.962 \\ 
Filling Factor & 0.882 & 0.0068 & 0.054 & -1.564 & 0.989 & 0.692 & 0.958 \\ 
Stray Light DS & 0.907 & 0.0005 & 0.009 & -3.414 & 1.022 & 0.750 & 0.960 \\ \bottomrule
\end{tabularx}
\caption{Performance metrics of the convolutional model ensemble.}
\label{p_metrics_ens}
\end{table*}

The data for training and validation were collected in such a way that they contained both the quiet and active areas of the Sun. The reference values $x_{true}$ refer to the result of Stokes profiles inversion by the ME codes HAO "MERLIN"~\citep{csac2006hinode2} (see Fig.~(\ref{true_params})). The metrics are represented in Tab.(\ref{p_metrics}). It can be seen that all the prediction parameters are covered by confidence intervals. The two-dimensional color maps of the 11 reconstructed parameters are visualized in Fig.~(\ref{params}) and comparison of these with the reference values can be seen in Fig.~(\ref{comparison}).

\begin{figure*}
\centering
\includegraphics[width=1\textwidth]{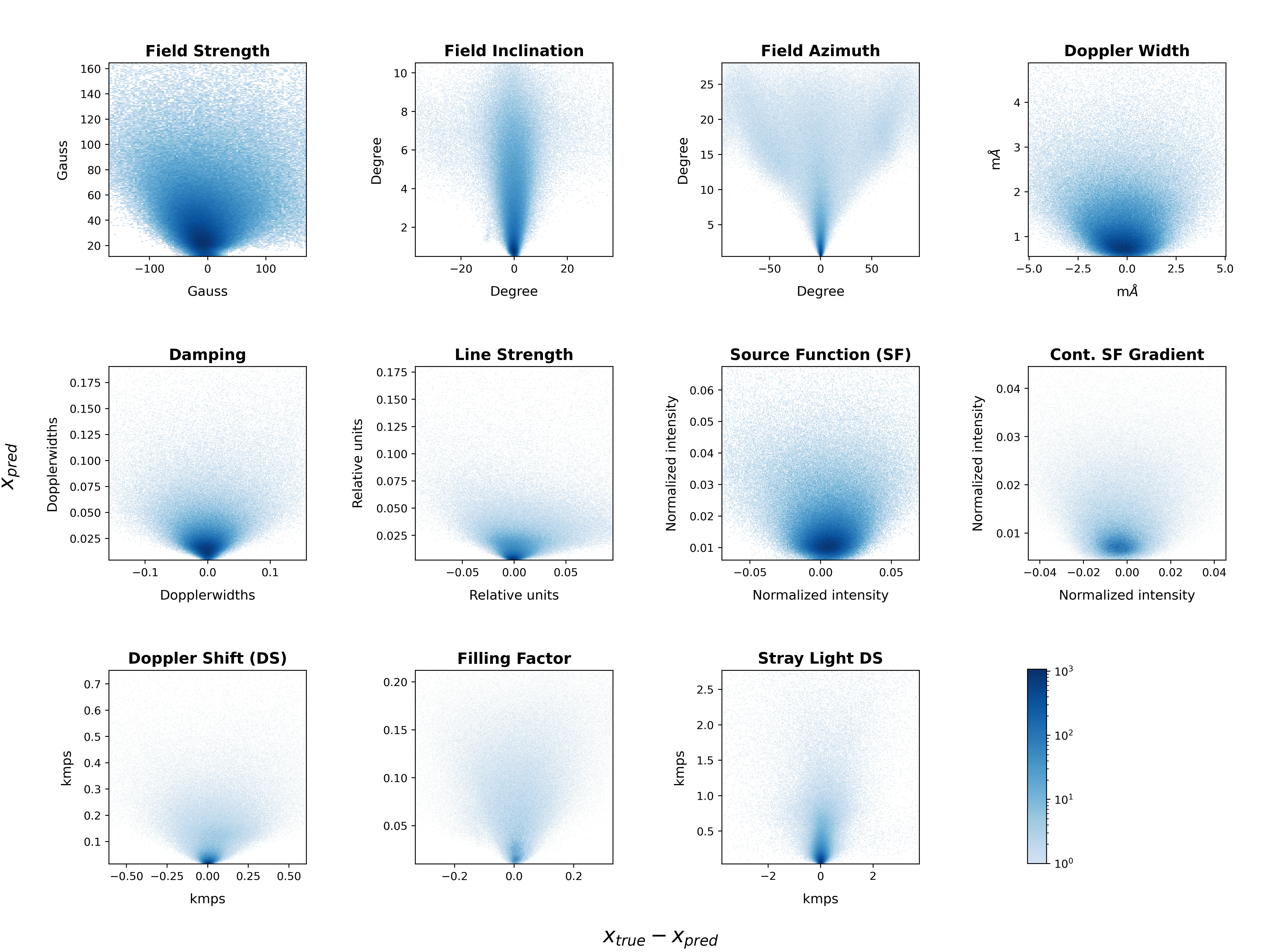}
\caption{Dependence of standard deviations on the difference between predictions and true values. Each figure corresponds to one of the atmospheric parameters. Color intensity indicates the frequency of the values encountered.}
\label{unc_1}
\end{figure*}

To estimate the performance of the method we also suggest plotting several dependencies. The scatter plots of the the standard deviation $\sigma_{pred}$ on the difference $x_{true} - x_{pred}$ are presented in Fig.~(\ref{unc_1}). Color intensity indicates the frequency of the values encountered. As one can see, the areas with the highest concentration of points are located near the zero error line. However, in the edges of the parameters, there are significant deviations of predictions from the reference values. The reason for this behavior could be the small number or complete absence of samples with such parameter values in training data.

\begin{figure*} 
\centering
\includegraphics[width=1\textwidth]{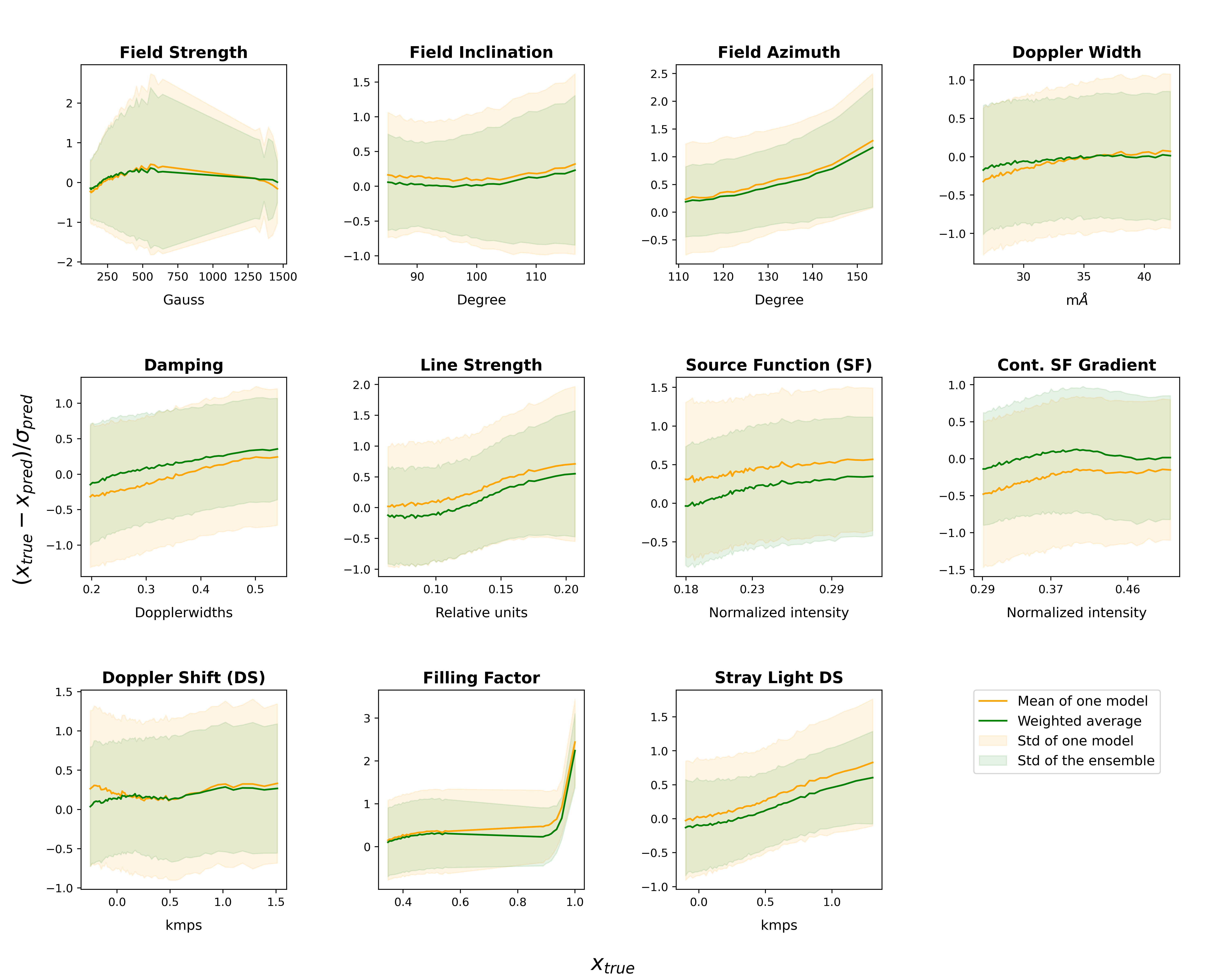}
\caption{Analysis of the difference between predictions and true values divided by standard deviation as a function of true values. Being segmented, these data were fitted by a normal distribution, then the dependence of the parameters of this approximation on the true data was constructed. Each figure corresponds to one of the atmospheric parameters. In case of one single model, mean values are marked by solid orange lines and the confidence intervals are by translucent orange areas. The average curves obtained from an ensemble are marked by solid green lines, the corresponding confidence intervals by translucent green areas.}
\label{unc_2}
\end{figure*}

For further analysis, dependencies of the ratio $(x_{true} - x_{pred})/ \sigma_{pred}$ on the true values $x_{true}$ in case of each parameter were divided into approximately $500$ segments, and then each segment was fitted by a normal distribution. These fitting curves can be seen in Fig.~(\ref{unc_2}). In the ideal case, the mean values have to be close to $0$, while the standard deviations have to be in the range $\{-1;1\}$. It can be seen that for some parameters there are deviations from the ideal scenario. Regions known as the most difficult to reconstruct (such as low Field Strength) have some over- or under-estimation of uncertainty, but not more than 30\%.

\begin{figure*}
\centering
\includegraphics[width=1\textwidth]{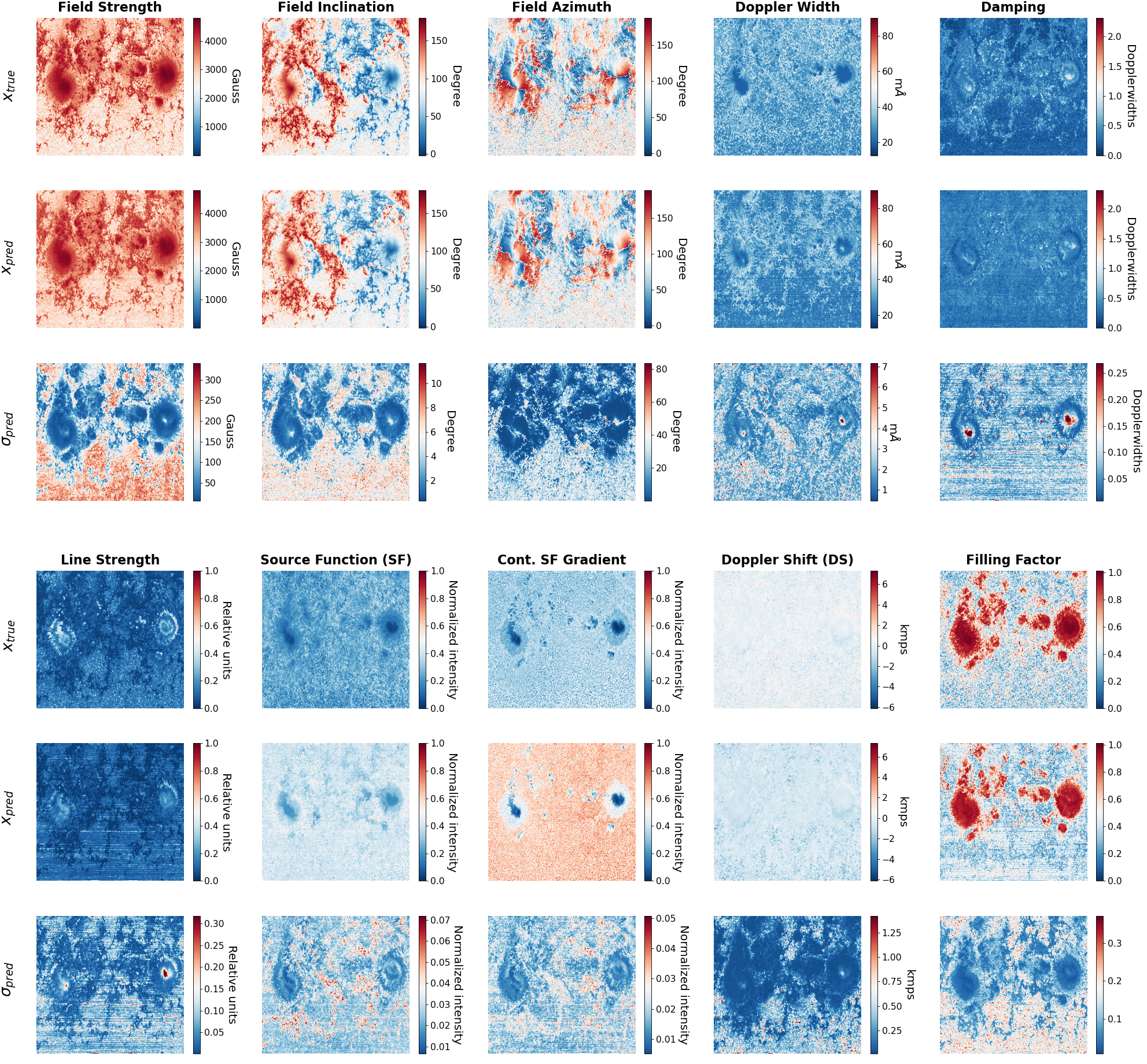}
\caption{Two-dimensional color maps of the results of Stokes profiles inversion by the ME codes HAO "MERLIN”, model predictions made on real data of Stokes profiles and results of uncertainty estimation in these predictions. Each column corresponds to one of the first 10 atmospheric parameters. For each parameter, the true values are shown in the first line, predictions in the second line and results of uncertainty estimation in the third line.}
\label{real_data}
\end{figure*}

In addition, an ensemble was created in order to improve the quality of predictions based on single convolutional models and using bagging. The performance metrics of the ensemble can be seen in Tab. (\ref{p_metrics_ens}). A comparison of the results obtained by one model and the ensemble of models is in Fig.~(\ref{unc_2}). As can be seen, the use of ensembles leads to a reduction of prediction variance and smoothing of the results. 

\label{sect:realdata}
\section{Testing on the real Hinode observations}
The model was also tested on the real data, Hinode/SOT/SP observations~\citep{csac2006hinode} collected on 26 September 2014 for NOAA 12172 active region, that is 12 times smaller than the size of our train data set. Comparison between the predictions made from the real Stokes parameters and the test data is on the Fig.~(\ref{real_data}). It can be seen that the new approach infers atmospheric parameters with an accuracy comparable to the ME inversion technique. 

It could also be noted that the model provides physically adequate uncertainties. For example, regions of weak magnetic field (where Stokes Q, U and V parameters have low amplitudes, and, thus, bad signal-to-noise ratio) correspond to predictions with larger uncertainties for the magnetic field vector. At the same time, pixels with strong magnetic field (where all 4 Stokes parameters suffer from low signal-to-noise ratio cause lack of light intensity) match predictions with big uncertainties for almost all of the parameters.

In some of the more noisy regions of the Sun, the model predicts less accurate and over confident results, thus leaving space for improvement of the model as well as the method of generating synthetic spectra. We assume that performance quality and robustness to out-of-distribution samples can be improved if the synthetic generation algorithm is upgraded, since synthetic profiles are usually symmetrical, while distribution of the real data could be uneven and asymmetrical. 

\label{sect:discussion}
\section{Conclusion and discussion} 
Machine learning, in particular the neural network approach to Stokes profile inversion, is gaining popularity due to computational efficiency, but physical models often require not only point estimates, but also errors. In this paper we provide a novel neural network architecture for inferring solar atmospheric parameters, together with their predictive uncertainties by modifying model architecture and loss function. The method was tested on the 11 atmospheric parameters: three components of the magnetic field vector, three line parameters, two intensity parameters, Doppler shift of the line, stray light Doppler shift and magnetic filling factor. The several performance metrics were calculated (R$^2$, MSE, MAE, NLPD, nRMSE, PICP$_{68}$, and PICP$_{95}$) in case of synthetic data, as well as real Hinode observations collected on 26 September 2014 for NOAA 12172 active region. The maps of Stokes profiles inversion by Milne-Eddington codes HAO "MERLIN" were taken as a ground truth, and about a minute was required for the model to make predictions of one map of 512$\times$873 pixels. Analysis showed that the proposed model represents a reliable method compared with classical methods for solving the inverse problem. In addition, it has been shown that the smoothness and the accuracy of results, and the width of the uncertainty intervals can be improved by ensembles. On a reduced set of observations, we show that the proposed method provides reasonable results and thus, can be used to improve theoretical calculations and provide a starting point for more precise methods, thus making it possible to reduce the total computation time. Although synthetic spectra are symmetric, which is unusual for the real data, the model trained on synthetic spectra showed the ability to generalize even in the case of real observations. 

The method proposed can be used for analysis in various fields of astrophysics~\citep{podladchikova2022maximal, okamoto2009prominence}: in the analysis of the solar cycle and prediction of coronal mass ejection, in the analysis of the solar atmosphere itself, for example, to study the spatial distribution of parameters or of the processes in the solar atmosphere such plasma convection, and open prospects for future studies. Further analysis raises the question of the credibility of the results obtained by the network. The model described is a simple and scalable method for quantifying uncertainty. Since it necessitates only a modification of architecture (doubling the number of output layers and changing the loss function), it requires as much learning time as a network that is not modified. 

\begin{table}
\renewcommand\tabcolsep{4pt}
\begin{tabularx}{1\textwidth}{llll}
\toprule
\textbf{Reference} &  
\textbf{Network architecture} &  
\textbf{Input data} &  
\textbf{Output data}
\\ \midrule

Carroll and Staude,~\citeyear{carroll2001the}
& MLP blocks       
& Synthetic   
& 
\begin{tabular}[t]{@{}l}
MF, V, FF \\ [1.5ex]
and other
\end{tabular}
\\ \hline

\begin{tabular}[t]{@{}l}
Socas-Navarro,~\citeyear{socas-navarro2005strategies} 
\end{tabular}
& 
\begin{tabular}[t]{@{}l}
MLP blocks 
\end{tabular}
& 
\begin{tabular}[t]{@{}l}
    Synthetic and real\\ [1.5ex]
    from High Altitude\\ [1.5ex]
    Observatory (HAO)
\end{tabular}
&
\begin{tabular}[t]{@{}l}
    MF, DW, \\ [1.5ex]
    SF, FF, LD \\ [1.5ex]
    and other
\end{tabular}  
\\ \hline

Ramos et al.,~\citeyear{ramos2007bayesian} 
& 
\begin{tabular}[t]{@{}l}
Bayesian \\ [1.5ex]
inference 
\end{tabular}        
& Synthetic    
& MF 
\\ \hline

Ramos et al.,~\citeyear{ramos2017inference} 
& 
\begin{tabular}[t]{@{}l}
Variational\\ [1.5ex]
inference 
\end{tabular}        
& 
\begin{tabular}[t]{@{}l}
    Real from \\ [1.5ex] 
    Swedish Solar \\ [1.5ex] 
    Telescope (SST)
\end{tabular}
&
\begin{tabular}[t]{@{}l}
   MF 
\end{tabular}
\\ \hline

Ramos and Baso,~\citeyear{ramos2019stokes} 
&        
\begin{tabular}[t]{@{}l}
    CNN with 2D \\ [1.5ex]
    convolutions
\end{tabular}
& Synthetic    
&
\begin{tabular}[t]{@{}l}
    MF, V, T \\ [1.5ex]
    and other
\end{tabular}      
\\ \hline

Sainz Dalda et al.,~\citeyear{sainzdalda2019recovering}
& 
\begin{tabular}[t]{@{}l}
    MLP blocks 
\end{tabular} 
& Synthetic
&
\begin{tabular}[t]{@{}l}
    V, T and other
\end{tabular}  
\\ \hline

Li et al.,~\citeyear{li2019mcmc} 
& 
\begin{tabular}[t]{@{}l}
Bayesian \\ [1.5ex]
inference 
\end{tabular} 
& Synthetic    
&
\begin{tabular}[t]{@{}l}
     MF, V, DW, \\ [1.5ex]
     SF + SFG, \\ [1.5ex]
     LD and other
\end{tabular}
\\ \hline

Liu et al.,~\citeyear{liu2020inferring} 
&        
\begin{tabular}[t]{@{}l}
    CNN with 1D \\ [1.5ex]
    convolutions
\end{tabular}       
& 
\begin{tabular}[t]{@{}l}
    Real from \\ [1.5ex]
    Goode Solar \\ [1.5ex] 
    Telescope (GST)
\end{tabular}      
&
\begin{tabular}[t]{@{}l}
    MF
\end{tabular}
\\ \hline

Mili{\'c} and Gafeira,~\citeyear{milic2020mimicking}
&     
\begin{tabular}[t]{@{}l}
    CNN with 1D \\ [1.5ex]
    convolutions
\end{tabular}     
& Synthetic      
&
\begin{tabular}[t]{@{}l}
    MF, V, T
\end{tabular}       
\\ \hline

Gafeira,~\citeyear{gafeira2021machine} 
& 
\begin{tabular}[t]{@{}l}
    Ensemble of \\ [1.5ex]
    CNNs with 1D \\ [1.5ex]
    convolutions
\end{tabular}        
& 
\begin{tabular}[t]{@{}l}
    Synthetic and \\ [1.5ex]
    real from Gregor \\ [1.5ex]
    telescope     
\end{tabular}
&
\begin{tabular}[t]{@{}l}
    MF, V, T
\end{tabular}       
\\ \hline

Guo et al.,~\citeyear{guo2021a}  
& 
\begin{tabular}[t]{@{}l}
    CNN with 2D \\ [1.5ex]
    convolutions   
\end{tabular}    
& 
\begin{tabular}[t]{@{}l}
    Real from \\ [1.5ex]
    Hinode telescope      
\end{tabular}  
&
\begin{tabular}[t]{@{}l}
    MF
\end{tabular}
\\ \hline
 
Higgins et al.,~\citeyear{higgins2021fast}  
&     
\begin{tabular}[t]{@{}l}
    CNN (U-Net), \\ [1.5ex]
    treating the problem \\ [1.5ex]
    as a regression \\ [1.5ex]
    by classification
\end{tabular}       
&
\begin{tabular}[t]{@{}l}
    Real from Solar \\ [1.5ex]
    Dynamics \\ [1.5ex]
    Observatory (SDO)
\end{tabular}
& 
\begin{tabular}[t]{@{}l}
    MF, DW, \\ [1.5ex]
    SF + SFG,\\ [1.5ex]
    and other
\end{tabular}
\\ \hline

Baso et al.,~\citeyear{baso2021bayesian}
& 
\begin{tabular}[t]{@{}l}
    Variational \\ [1.5ex]
    inference
\end{tabular}
&
\begin{tabular}[t]{@{}l}
    Synthetic and real \\ [1.5ex]
    from Swedish Solar\\ [1.5ex]
    Telescope (SST)
\end{tabular}  
&
\begin{tabular}[t]{@{}l}
    V, T, DW, \\ [1.5ex]
    SF + SFG\\ [1.5ex] 
    and other
\end{tabular}
\\ \hline

Knyazeva et al.,~\citeyear{knyazeva2022multi}  
& MLP blocks        
& Synthetic    
& 
\begin{tabular}[t]{@{}l}
    MF, DW, FF, \\ [1.5ex]
    SF + SFG, \\ [1.5ex]
    LD and other
\end{tabular} 
\\ \hline

Higgins et al.,~\citeyear{higgins2022synthia}  
&     
\begin{tabular}[t]{@{}l}
    CNN (U-Net), \\ [1.5ex]
    treating the problem \\ [1.5ex]
    as a regression \\ [1.5ex]
    by classification
\end{tabular}       
& Synthetic    
& 
\begin{tabular}[t]{@{}l}
    MF, FF \\ [1.5ex]
    and other 
\end{tabular}
\\ \hline

Present work 
&     
\begin{tabular}[t]{@{}l}
    Ensemble of CNNs \\ [1.5ex]
    with 1D convolutions \\ [1.5ex]
    and uncertainty \\ [1.5ex]
    quantification
\end{tabular}       
&  
\begin{tabular}[t]{@{}l}
    Synthetic and \\ [1.5ex]
    real from Hinode \\ [1.5ex]
    telescope
\end{tabular}

& 
\begin{tabular}[t]{@{}l}
    MF, DW, FF, \\ [1.5ex]
    SF + SFG,  \\ [1.5ex]
    LD and other
\end{tabular} 
\\ \bottomrule
\end{tabularx}
\caption{Overview of neural network studies in Stocks inversion problem. The following abbreviations have been used in this table: 
Magnetic Field (MF), 
Velocity (V), 
Temperature (T), 
Source Function and its Gradient (SF and SFG), 
Doppler Width (DW),
Filling Factor (FF) and 
Line Damping (LD).
}
\label{review}
\end{table}

\label{sect:availability}
\section*{Data Availability}
In the current study, we used a collection of the Level 1 calibrated Stokes spectra (comprised by images stored in FITS format) and collection of the Level 2 data sets (obtained from the MERLIN spectral line inversion of the Level 1 calibrated spectra) produced by the Spectropolarimeter (SP) on board the Hinode, since its launch in 2006. Hinode is a Japanese mission, developed and launched by ISAS/JAXA, with NAOJ as a domestic partner and NASA and STFC (UK) as international partners. It is operated by these agencies in co-operation with ESA and NSC (Norway). The Hinode has an open data policy, allowing anyone access to the data and data
products. Level 1 and 2 data are available by following the \href{https://csac.hao.ucar.edu/sp_data.php}{data link}.

\label{sect:acks}
\section*{Acknowledgements}
Denis Derkach, Lukia Mistryukova,
Aleksandr Khizhik and Mikhail Hushchyn are grateful to the HSE basic research program. This research was supported in part through computational resources of HPC facilities at HSE University (Kostenetskiy, Chulkevich, and Kozyrev,~\citeyear{kostenetskiy2021hpc}).

\label{sect:bib}
\bibliographystyle{spr-mp-sola}
\bibliography{InverseSolar}

\begin{thebibliography}{41}
\ifx\bisbn     \undefined \def\bisbn  #1{ISBN #1}\fi
\ifx\binits    \undefined \def\binits#1{#1}\fi
\ifx\bauthor   \undefined \def\bauthor#1{#1}\fi
\ifx\batitle   \undefined \def\batitle#1{#1}\fi
\ifx\bjtitle   \undefined \def\bjtitle#1{\textit{#1}}\fi
\ifx\bvolume   \undefined \def\bvolume#1{\textbf{#1}}\fi
\ifx\byear     \undefined \def\byear#1{#1}\fi
\ifx\bissue    \undefined \def\bissue#1{#1}\fi
\ifx\bfpage    \undefined \def\bfpage#1{#1}\fi
\ifx\blpage    \undefined \def\blpage #1{#1}\fi
\ifx\burl      \undefined \def\burl#1{\textsf{#1}}\fi
\ifx\href      \undefined \def\href#1#2{\textsf{#2}}\fi
\ifx\betal     \undefined \def\betal{\textit{et al.}}\fi
\ifx\bctitle   \undefined \def\bctitle#1{#1}\fi
\ifx\beditor   \undefined \def\beditor#1{#1}\fi
\ifx\bbtitle   \undefined \def\bbtitle#1{\textit{#1}}\fi
\ifx\bedition  \undefined \def\bedition#1{#1}\fi
\ifx\bseriesno \undefined \def\bseriesno#1{\textbf{#1}}\fi
\ifx\blocation \undefined \def\blocation#1{#1}\fi
\ifx\bsertitle \undefined \def\bsertitle#1{\textit{#1}}\fi
\ifx\bsnm      \undefined \def\bsnm#1{#1}\fi
\ifx\bsuffix   \undefined \def\bsuffix#1{#1}\fi
\ifx\bparticle \undefined \def\bparticle#1{#1}\fi
\ifx\barticle  \undefined \def\barticle#1{}\fi
\ifx\binstitute  \undefined \def\binstitute#1{#1}\fi
\ifx\bpublisher  \undefined \def\bpublisher#1{#1}\fi
\ifx\doiurl    \undefined
  \def\doiurl#1{\href{http://dx.doi.org/#1}{\textsf{DOI}}}\fi
\ifx\arxivurl  \undefined
  \def\arxivurl#1{\href{http://arxiv.org/abs/#1}{\textsf{arXiv}}}\fi
\ifx\adsurl    \undefined
  \def\adsurl#1{\href{http://adsabs.harvard.edu/abs/#1}{\textsf{ADS}}}\fi
\ifx\botherref \undefined \def\botherref#1{}\fi
\ifx\url       \undefined \def\url#1{\textsf{#1}}\fi
\ifx\bchapter  \undefined \def\bchapter#1{}\fi
\ifx\bbook     \undefined \def\bbook#1{}\fi
\ifx\bcomment  \undefined \def\bcomment#1{#1}\fi
\ifx\oauthor   \undefined \def\oauthor#1{#1}\fi
\ifx\citeauthoryear \undefined\def \citeauthoryear#1{#1}\fi
\ifx\endbibitem\undefined \def\endbibitem{}\fi
\ifx\bconflocation  \undefined \def\bconflocation#1{#1} \fi

\bibitem[\protect\citeauthoryear{Baso, Ramos, and de~la
  Cruz~Rodríguez}{2022}]{baso2021bayesian}
\begin{botherref}
\oauthor{\bsnm{Baso}, \binits{C.J.D.}},
\oauthor{\bsnm{Ramos}, \binits{A.A.}},
\oauthor{\bparticle{de~la} \bsnm{Cruz~Rodríguez}, \binits{J.}}:
2022,
Bayesian stokes inversion with normalizing flows.
\textit{\aap}.
\end{botherref}
\endbibitem

\bibitem[\protect\citeauthoryear{Blundell
  \textit{et~al.}}{2015}]{blundell2015weight}
\begin{botherref}
\oauthor{\bsnm{Blundell}, \binits{C.}},
\oauthor{\bsnm{Cornebise}, \binits{J.}},
\oauthor{\bsnm{Kavukcuoglu}, \binits{K.}},
\oauthor{\bsnm{Wierstra}, \binits{D.}}:
2015,
\textit{Weight uncertainty in neural networks},
arXiv.
\end{botherref}
\endbibitem

\bibitem[\protect\citeauthoryear{Borrero
  \textit{et~al.}}{2011}]{borrero2011vfisv}
\begin{barticle}
\bauthor{\bsnm{Borrero}, \binits{J.M.}},
\bauthor{\bsnm{Tomczyk}, \binits{S.}},
\bauthor{\bsnm{Kubo}, \binits{M.}},
\bauthor{\bsnm{Socas-Navarro}, \binits{H.}},
\bauthor{\bsnm{Schou}, \binits{J.}}, \betal:
\byear{2011},
\batitle{{VFISV: Very Fast Inversion of the Stokes Vector for the Helioseismic
  and Magnetic Imager}}.
\bjtitle{\solphys}
\bvolume{273}(\bissue{1}),
\bfpage{267}.
\end{barticle}
\endbibitem

\bibitem[\protect\citeauthoryear{Carroll and Staude}{2001}]{carroll2001the}
\begin{barticle}
\bauthor{\bsnm{Carroll}, \binits{T.A.}},
\bauthor{\bsnm{Staude}, \binits{J.}}:
\byear{2001},
\batitle{{The inversion of Stokes profiles with artificial neural networks}}.
\bjtitle{\aap}
\bvolume{378},
\bfpage{316}.
\end{barticle}
\endbibitem

\bibitem[\protect\citeauthoryear{{Community Spectropolarimetric Analysis Center
  (CSAC)}}{2006a}]{csac2006hinode}
\begin{botherref}
\oauthor{\bsnm{{Community Spectropolarimetric Analysis Center (CSAC)}}}:
2006a,
\textit{Hinode-spectropolarimeter (sp) level 1 (calibrated) full stokes data},
UCAR/NCAR - HAO/Community Spectropolarimetric Analysis Center.
\end{botherref}
\endbibitem

\bibitem[\protect\citeauthoryear{{Community Spectropolarimetric Analysis Center
  (CSAC)}}{2006b}]{csac2006hinode2}
\begin{botherref}
\oauthor{\bsnm{{Community Spectropolarimetric Analysis Center (CSAC)}}}:
2006b,
\textit{Hinode-spectropolarimeter (sp) level 2 (vector magnetic field) spectral
  line inversions},
UCAR/NCAR - HAO/Community Spectropolarimetric Analysis Center.
\end{botherref}
\endbibitem

\bibitem[\protect\citeauthoryear{Dalda
  \textit{et~al.}}{2019}]{sainzdalda2019recovering}
\begin{barticle}
\bauthor{\bsnm{Dalda}, \binits{A.S.}},
\bauthor{},
\bauthor{\bparticle{de~la} \bsnm{Cruz~Rodríguez}, \binits{J.}},
\bauthor{\bsnm{Pontieu}, \binits{B.D.}},
\bauthor{\bsnm{Go{\v s}i{\'c}}, \binits{M.}}:
\byear{2019},
\batitle{Recovering thermodynamics from spectral profiles observed by {IRIS}: A
  machine and deep learning approach}.
\bjtitle{Astrophys. J. Lett.}
\bvolume{875}(\bissue{2}),
\bfpage{L18}.
\end{barticle}
\endbibitem

\bibitem[\protect\citeauthoryear{del Toro~Iniesta and
  Ruiz~Cobo}{2016}]{DelToroIniesta2016}
\begin{botherref}
\oauthor{\bparticle{del} \bsnm{Toro~Iniesta}, \binits{J.C.}},
\oauthor{\bsnm{Ruiz~Cobo}, \binits{B.}}:
2016,
Inversion of the radiative transfer equation for polarized light.
\textit{Liv. Rev. Sol. Phys.}
\textbf{13}(1).
\end{botherref}
\endbibitem

\bibitem[\protect\citeauthoryear{Gafeira
  \textit{et~al.}}{2021}]{gafeira2021machine}
\begin{barticle}
\bauthor{\bsnm{Gafeira}, \binits{R.}},
\bauthor{\bsnm{Su{\'a}rez}, \binits{D.O.}},
\bauthor{\bsnm{Mili{\'c}}, \binits{I.}},
\bauthor{\bsnm{Noda}, \binits{C.Q.}},
\bauthor{\bsnm{Cobo}, \binits{B.R.}}, \betal:
\byear{2021},
\batitle{Machine learning initialization to accelerate stokes profile
  inversions}.
\bjtitle{\aap}
\bvolume{651},
\bfpage{A31}.
\end{barticle}
\endbibitem

\bibitem[\protect\citeauthoryear{Gawlikowski
  \textit{et~al.}}{2021}]{gawlikowski2021a}
\begin{botherref}
\oauthor{\bsnm{Gawlikowski}, \binits{J.}},
\oauthor{\bsnm{Tassi}, \binits{C.R.N.}},
\oauthor{\bsnm{Ali}, \binits{M.}},
\oauthor{\bsnm{Lee}, \binits{J.}},
\oauthor{\bsnm{Humt}, \binits{M.}}, et al.:
2021,
A survey of uncertainty in deep neural networks.
\textit{arXiv preprint arXiv:2107.03342}.
\end{botherref}
\endbibitem

\bibitem[\protect\citeauthoryear{Ghahramani}{2015}]{ghahramani2015probabilistic}
\begin{barticle}
\bauthor{\bsnm{Ghahramani}, \binits{Z.}}:
\byear{2015},
\batitle{Probabilistic machine learning and artificial intelligence}.
\bjtitle{Nature}
\bvolume{521}(\bissue{7553}),
\bfpage{452}.
\end{barticle}
\endbibitem

\bibitem[\protect\citeauthoryear{Grossmann-Doerth, Knölker, and
  Schuessler}{1994}]{grossmann1994thedeep}
\begin{barticle}
\bauthor{\bsnm{Grossmann-Doerth}, \binits{U.}},
\bauthor{\bsnm{Knölker}, \binits{M.}},
\bauthor{\bsnm{Schuessler}, \binits{M.}}:
\byear{1994},
\batitle{The deep layers of solar magnetic elements}.
\bjtitle{Astronomy and Astrophysics}
\bvolume{285},
\bfpage{648}.
\end{barticle}
\endbibitem

\bibitem[\protect\citeauthoryear{Guo \textit{et~al.}}{2021}]{guo2021a}
\begin{barticle}
\bauthor{\bsnm{Guo}, \binits{J.}},
\bauthor{\bsnm{Bai}, \binits{X.}},
\bauthor{\bsnm{Liu}, \binits{H.}},
\bauthor{\bsnm{Yang}, \binits{X.}},
\bauthor{\bsnm{Deng}, \binits{Y.}}, \betal:
\byear{2021},
\batitle{A nonlinear solar magnetic field calibration method for the
  filter-based magnetograph by the residual network}.
\bjtitle{\aap}
\bvolume{646},
\bfpage{A41}.
\end{barticle}
\endbibitem

\bibitem[\protect\citeauthoryear{Higgins
  \textit{et~al.}}{2021}]{higgins2021fast}
\begin{barticle}
\bauthor{\bsnm{Higgins}, \binits{R.E.L.}},
\bauthor{\bsnm{Fouhey}, \binits{D.F.}},
\bauthor{\bsnm{Zhang}, \binits{D.}},
\bauthor{\bsnm{Antiochos}, \binits{S.K.}},
\bauthor{\bsnm{Barnes}, \binits{G.}}, \betal:
\byear{2021},
\batitle{Fast and accurate emulation of the {SDO}/{HMI} stokes inversion with
  uncertainty quantification}.
\bjtitle{The Astrophysical Journal}
\bvolume{911}(\bissue{2}),
\bfpage{130}.
\end{barticle}
\endbibitem

\bibitem[\protect\citeauthoryear{Higgins
  \textit{et~al.}}{2022}]{higgins2022synthia}
\begin{barticle}
\bauthor{\bsnm{Higgins}, \binits{R.E.L.}},
\bauthor{\bsnm{Fouhey}, \binits{D.F.}},
\bauthor{\bsnm{Antiochos}, \binits{S.K.}},
\bauthor{\bsnm{Barnes}, \binits{G.}},
\bauthor{\bsnm{Cheung}, \binits{M.C.M.}}, \betal:
\byear{2022},
\batitle{{SynthIA}: A synthetic inversion approximation for the stokes vector
  fusing {SDO} and hinode into a virtual observatory}.
\bjtitle{The Astrophysical Journal Supplement Series}
\bvolume{259}(\bissue{1}),
\bfpage{24}.
\end{barticle}
\endbibitem

\bibitem[\protect\citeauthoryear{Knyazeva
  \textit{et~al.}}{2022}]{knyazeva2022multi}
\begin{bbook}
\bauthor{\bsnm{Knyazeva}, \binits{I.}},
\bauthor{\bsnm{Plotnikov}, \binits{A.}},
\bauthor{\bsnm{Medvedeva}, \binits{T.}},
\bauthor{\bsnm{Makarenko}, \binits{N.}}:
\byear{2022},
\bbtitle{Multi-output deep learning framework for solar atmospheric parameters
  inferring from stokes profiles}.
\end{bbook}
\endbibitem

\bibitem[\protect\citeauthoryear{Kostenetskiy, Chulkevich, and
  Kozyrev}{2021}]{kostenetskiy2021hpc}
\begin{barticle}
\bauthor{\bsnm{Kostenetskiy}, \binits{P.S.}},
\bauthor{\bsnm{Chulkevich}, \binits{R.A.}},
\bauthor{\bsnm{Kozyrev}, \binits{V.I.}}:
\byear{2021},
\batitle{{HPC} resources of the higher school of economics}.
\bjtitle{Journal of Physics: Conference Series}
\bvolume{1740}(\bissue{1}),
\bfpage{012050}.
\end{barticle}
\endbibitem

\bibitem[\protect\citeauthoryear{Krzywinski and
  Altman}{2013}]{krzywinski2013power}
\begin{barticle}
\bauthor{\bsnm{Krzywinski}, \binits{M.}},
\bauthor{\bsnm{Altman}, \binits{N.}}:
\byear{2013},
\batitle{Power and sample size}.
\bjtitle{Nature Methods}
\bvolume{10}(\bissue{12}),
\bfpage{1139}.
\end{barticle}
\endbibitem

\bibitem[\protect\citeauthoryear{Kuckein
  \textit{et~al.}}{2021}]{kuckein2021multiple}
\begin{barticle}
\bauthor{\bsnm{Kuckein}, \binits{C.}},
\bauthor{\bsnm{Balthasar}, \binits{H.}},
\bauthor{\bsnm{Noda}, \binits{C.Q.}},
\bauthor{\bsnm{Diercke}, \binits{A.}},
\bauthor{\bsnm{Arjona}, \binits{J.C.T.}}, \betal:
\byear{2021},
\batitle{Multiple stokes i inversions for inferring magnetic fields in the
  spectral range around cr 782 $\&$}.
\bjtitle{\aap}
\bvolume{653},
\bfpage{A165}.
\end{barticle}
\endbibitem

\bibitem[\protect\citeauthoryear{Lakshminarayanan, Pritzel, and
  Blundell}{2017}]{lakshminarayanan2017simple}
\begin{bchapter}
\bauthor{\bsnm{Lakshminarayanan}, \binits{B.}},
\bauthor{\bsnm{Pritzel}, \binits{A.}},
\bauthor{\bsnm{Blundell}, \binits{C.}}:
\byear{2017},
\bctitle{Simple and scalable predictive uncertainty estimation using deep
  ensembles}.
In: \bbtitle{Advances in Neural Information Processing Systems},
\bsertitle{Nips'17}.
\end{bchapter}
\endbibitem

\bibitem[\protect\citeauthoryear{Landi~Degl'Innocenti and
  Landolfi}{2004}]{degl2004polarization}
\begin{botherref}
\oauthor{\bsnm{Landi~Degl'Innocenti}, \binits{E.}},
\oauthor{\bsnm{Landolfi}, \binits{M.}}:
2004,
\textit{{Polarization in Spectral Lines}},
Springer Dordrecht.
\end{botherref}
\endbibitem

\bibitem[\protect\citeauthoryear{Leka \textit{et~al.}}{2022}]{leka2022on}
\begin{botherref}
\oauthor{\bsnm{Leka}, \binits{K.D.}},
\oauthor{\bsnm{Wagner}, \binits{E.L.}},
\oauthor{\bsnm{Griñón-Marín}, \binits{A.B.}},
\oauthor{\bsnm{Bommier}, \binits{V.}},
\oauthor{\bsnm{Higgins}, \binits{R.}}:
2022,
\textit{On identifying and mitigating bias in inferred measurements for solar
  vector magnetic field data},
arXiv.
\end{botherref}
\endbibitem

\bibitem[\protect\citeauthoryear{Li \textit{et~al.}}{2019}]{li2019mcmc}
\begin{barticle}
\bauthor{\bsnm{Li}, \binits{H.}},
\bauthor{\bsnm{Xu}, \binits{Z.}},
\bauthor{\bsnm{Qu}, \binits{Z.}},
\bauthor{\bsnm{Sun}, \binits{L.}}:
\byear{2019},
\batitle{{MCMC} inversion of stokes profiles}.
\bjtitle{The Astrophysical Journal}
\bvolume{875}(\bissue{2}),
\bfpage{127}.
\end{barticle}
\endbibitem

\bibitem[\protect\citeauthoryear{Li \textit{et~al.}}{2022}]{li2022tic}
\begin{barticle}
\bauthor{\bsnm{Li}, \binits{H.}},
\bauthor{\bparticle{del} \bsnm{Pino~Alem{\'{a}}n}, \binits{T.}},
\bauthor{\bsnm{Bueno}, \binits{J.T.}},
\bauthor{\bsnm{Casini}, \binits{R.}}:
\byear{2022},
\batitle{{TIC}: A stokes inversion code for scattering polarization with
  partial frequency redistribution and arbitrary magnetic fields}.
\bjtitle{The Astrophysical Journal}
\bvolume{933}(\bissue{2}),
\bfpage{145}.
\end{barticle}
\endbibitem

\bibitem[\protect\citeauthoryear{{Lites} \textit{et~al.}}{2007}]{lites2007a}
\begin{barticle}
\bauthor{\bsnm{{Lites}}, \binits{B.}},
\bauthor{\bsnm{{Casini}}, \binits{R.}},
\bauthor{\bsnm{{Garcia}}, \binits{J.}},
\bauthor{\bsnm{{Socas-Navarro}}, \binits{H.}}:
\byear{2007},
\batitle{{A suite of community tools for spectro-polarimetric analysis.}}
\bjtitle{MemSAIt}
\bvolume{78},
\bfpage{148}.
\end{barticle}
\endbibitem

\bibitem[\protect\citeauthoryear{Liu \textit{et~al.}}{2020}]{liu2020inferring}
\begin{barticle}
\bauthor{\bsnm{Liu}, \binits{H.}},
\bauthor{\bsnm{Xu}, \binits{Y.}},
\bauthor{\bsnm{Wang}, \binits{J.}},
\bauthor{\bsnm{Jing}, \binits{J.}},
\bauthor{\bsnm{Liu}, \binits{C.}}, \betal:
\byear{2020},
\batitle{Inferring vector magnetic fields from stokes profiles of {GST}/{NIRIS}
  using a convolutional neural network}.
\bjtitle{ApJ}
\bvolume{894}(\bissue{1}),
\bfpage{70}.
\end{barticle}
\endbibitem

\bibitem[\protect\citeauthoryear{Malinin and Gales}{}]{malinin2018predictive}
\begin{botherref}
\oauthor{\bsnm{Malinin}, \binits{A.}},
\oauthor{\bsnm{Gales}, \binits{M.}}:
Predictive uncertainty estimation via prior networks.
In: \textit{Advances in Neural Information Processing Systems},
Curran Associates, Inc..
\end{botherref}
\endbibitem

\bibitem[\protect\citeauthoryear{Mili{\'c} and
  Gafeira}{2020}]{milic2020mimicking}
\begin{barticle}
\bauthor{\bsnm{Mili{\'c}}, \binits{I.}},
\bauthor{\bsnm{Gafeira}, \binits{R.}}:
\byear{2020},
\batitle{Mimicking spectropolarimetric inversions using convolutional neural
  networks}.
\bjtitle{\aap}
\bvolume{644},
\bfpage{A129}.
\end{barticle}
\endbibitem

\bibitem[\protect\citeauthoryear{Okamoto
  \textit{et~al.}}{2009}]{okamoto2009prominence}
\begin{botherref}
\oauthor{\bsnm{Okamoto}, \binits{T.}},
\oauthor{\bsnm{Tsuneta}, \binits{S.}},
\oauthor{\bsnm{Lites}, \binits{B.}},
\oauthor{\bsnm{Kubo}, \binits{M.}},
\oauthor{\bsnm{Yokoyama}, \binits{T.}}, et al.:
2009,
Prominence formation associated with an emerging helical flux rope.
\textit{The Astrophysical Journal}
\textbf{697}.
\end{botherref}
\endbibitem

\bibitem[\protect\citeauthoryear{Pawitan}{2001}]{pawitan2001in}
\begin{botherref}
\oauthor{\bsnm{Pawitan}, \binits{Y.}}:
2001,
\textit{In all likelihood: Statistical modelling and inference using
  likelihood},
OUP Oxford.
\end{botherref}
\endbibitem

\bibitem[\protect\citeauthoryear{Podladchikova
  \textit{et~al.}}{2022}]{podladchikova2022maximal}
\begin{botherref}
\oauthor{\bsnm{Podladchikova}, \binits{T.}},
\oauthor{\bsnm{Jain}, \binits{S.}},
\oauthor{\bsnm{Veronig}, \binits{A.}},
\oauthor{\bsnm{Sutyrina}, \binits{O.}},
\oauthor{\bsnm{Dumbovic}, \binits{M.}}, et al.:
2022,
Maximal growth rate of the ascending phase of a sunspot cycle for predicting
  its amplitude.
\textit{\aap}.
\end{botherref}
\endbibitem

\bibitem[\protect\citeauthoryear{Quinonero-Candela
  \textit{et~al.}}{2005}]{quinonerocandela2005evaluating}
\begin{bchapter}
\bauthor{\bsnm{Quinonero-Candela}, \binits{J.}},
\bauthor{\bsnm{Rasmussen}, \binits{C.E.}},
\bauthor{\bsnm{Sinz}, \binits{F.}},
\bauthor{\bsnm{Bousquet}, \binits{O.}},
\bauthor{\bsnm{Sch{\"o}lkopf}, \binits{B.}}:
\byear{2005},
\bctitle{Evaluating predictive uncertainty challenge}.
In: \bbtitle{Machine Learning Challenges Workshop},
\bfpage{1}.
\bcomment{Springer}.
\end{bchapter}
\endbibitem

\bibitem[\protect\citeauthoryear{Ramos and Baso}{2019}]{ramos2019stokes}
\begin{botherref}
\oauthor{\bsnm{Ramos}, \binits{A.A.}},
\oauthor{\bsnm{Baso}, \binits{C.D.}}:
2019,
Stokes inversion based on convolutional neural networks.
\textit{\aap}
\textbf{626}(August).
\end{botherref}
\endbibitem

\bibitem[\protect\citeauthoryear{Ramos, Gonz{\'{a}}lez, and
  Rubi{\~{n}}o-Mart{\'{\i}}n}{2007}]{ramos2007bayesian}
\begin{barticle}
\bauthor{\bsnm{Ramos}, \binits{A.A.}},
\bauthor{\bsnm{Gonz{\'{a}}lez}, \binits{M.J.M.}},
\bauthor{\bsnm{Rubi{\~{n}}o-Mart{\'{\i}}n}, \binits{J.A.}}:
\byear{2007},
\batitle{Bayesian inversion of stokes profiles}.
\bjtitle{\aap}
\bvolume{476}(\bissue{2}),
\bfpage{959}.
\end{barticle}
\endbibitem

\bibitem[\protect\citeauthoryear{Ramos
  \textit{et~al.}}{2016}]{ramos2016inversion}
\begin{barticle}
\bauthor{\bsnm{Ramos}, \binits{A.A.}},
\bauthor{\bparticle{de~la} \bsnm{Cruz~Rodríguez}, \binits{J.}},
\bauthor{\bsnm{Gonzalez}, \binits{M.J.M.}},
\bauthor{\bsnm{Yabar}, \binits{A.P.}}:
\byear{2016},
\batitle{Inversion of stokes profiles with systematic effects}.
\bjtitle{\aap}
\bvolume{590},
\bfpage{A 87}.
\end{barticle}
\endbibitem

\bibitem[\protect\citeauthoryear{Ramos
  \textit{et~al.}}{2017}]{ramos2017inference}
\begin{barticle}
\bauthor{\bsnm{Ramos}, \binits{A.A.}},
\bauthor{\bparticle{de~la} \bsnm{Cruz~Rodríguez}, \binits{J.}},
\bauthor{\bsnm{Gonz{\'{a}}lez}, \binits{M.J.M.}},
\bauthor{\bsnm{Socas-Navarro}, \binits{H.}}:
\byear{2017},
\batitle{Inference of the chromospheric magnetic field orientation in the ca ii
  8542 a line fibrils}.
\bjtitle{\aap}
\bvolume{599},
\bfpage{A133}.
\end{barticle}
\endbibitem

\bibitem[\protect\citeauthoryear{Shorten and Khoshgoftaar}{2019}]{shorten2019a}
\begin{botherref}
\oauthor{\bsnm{Shorten}, \binits{C.}},
\oauthor{\bsnm{Khoshgoftaar}, \binits{T.M.}}:
2019,
A survey on image data augmentation for deep learning.
\textit{Journal of Big Data}
\textbf{6}(1).
\end{botherref}
\endbibitem

\bibitem[\protect\citeauthoryear{Shrestha and
  Solomatine}{2006}]{shrestha2006machine}
\begin{barticle}
\bauthor{\bsnm{Shrestha}, \binits{D.L.}},
\bauthor{\bsnm{Solomatine}, \binits{D.P.}}:
\byear{2006},
\batitle{Machine learning approaches for estimation of prediction interval for
  the model output}.
\bjtitle{Neural Networks}
\bvolume{19}(\bissue{2}),
\bfpage{225}.
\end{barticle}
\endbibitem

\bibitem[\protect\citeauthoryear{Socas-Navarro}{2005}]{socas-navarro2005strategies}
\begin{barticle}
\bauthor{\bsnm{Socas-Navarro}, \binits{H.}}:
\byear{2005},
\batitle{Strategies for spectral profile inversion using artificial neural
  networks}.
\bjtitle{ApJ}
\bvolume{621}(\bissue{1}),
\bfpage{545}.
\end{barticle}
\endbibitem

\bibitem[\protect\citeauthoryear{Unno}{1956}]{unno1956line}
\begin{barticle}
\bauthor{\bsnm{Unno}, \binits{W.}}:
\byear{1956},
\batitle{Line formation of a normal zeeman triplet}.
\bjtitle{Publications of the Astronomical Society of Japan}
\bvolume{8},
\bfpage{108}.
\end{barticle}
\endbibitem

\bibitem[\protect\citeauthoryear{Viticchi{\'{e}} and
  Almeida}{2011}]{viticchie2011asymmetries}
\begin{barticle}
\bauthor{\bsnm{Viticchi{\'{e}}}, \binits{B.}},
\bauthor{\bsnm{Almeida}, \binits{J.S.}}:
\byear{2011},
\batitle{Asymmetries of the stokes v profiles observed by hinode sot/sp in the
  quiet sun}.
\bjtitle{\aap}
\bvolume{530},
\bfpage{A14}.
\end{barticle}
\endbibitem

\end{thebibliography}

\end{article}
\end{document}